\documentclass[twocolumn,showpacs,preprintnumbers,amsmath,amssymb]{revtex4}
\input psfig.sty
\def \be {\begin{equation}}

\def \ee {\end{equation}}
\def \bea {\begin{eqnarray}}
\def \eea {\end{eqnarray}}

\usepackage{graphicx}
\usepackage{dcolumn}
\usepackage{bm}
\usepackage{epsfig}

\begin{document}

\title{Constraining the dark energy and smoothness parameter with type Ia Supernovae and Gamma-Ray Bursts}

\author{V. C. Busti$^{1}$} \email{vcbusti@astro.iag.usp.br}

\author{R. C. Santos$^{2}$} \email{cliviars@astro.iag.usp.br}

\author{J. A. S. Lima$^{1}$} \email{limajas@astro.iag.usp.br}
\vskip 1.5cm
\affiliation{$^{1}$Departamento de Astronomia, Universidade de S\~ao Paulo, S\~ao
Paulo, SP, Brasil}
\affiliation{$^{2}$Departamento de Ci\^encias Exatas e da Terra, Universidade Federal
de S\~ao Paulo,  Diadema, SP, Brasil}

\begin{abstract}
\noindent The existence of inhomogeneities in the observed Universe modifies the distance-redshift relations thereby affecting  the results of cosmological tests in comparison to the ones derived assuming spatially uniform models. 
By modeling  the inhomogeneities through a Zeldovich-Kantowski-Dyer-Roeder (ZKDR) approach which is phenomenologically characterized by a smoothness parameter $\alpha$, we rediscuss the constraints on the cosmic 
parameters based on type Ia Supernovae (SNe Ia) and Gamma-Ray Bursts (GRBs) data. The present analysis is  restricted to a flat $\Lambda$CDM model with the reasonable assumption that $\Lambda$ does not clump.  A $\chi^{2}$-analysis 
using 557 SNe Ia data from the Union2 Compilation Data (Amanullah {\it et al.} 2010) constrains the pair of  parameters ($\Omega_m, \alpha$) to  $\Omega_m=0.27_{-0.03}^{+0.08}$($2\sigma$) and $\alpha \geq 0.25$. A similar analysis 
based only on 59 Hymnium GRBs (Wei 2010) constrains the matter density parameter to be $\Omega_m= 0.35^{+0.62}_{-0.24}$ ($2\sigma$) while all values for the smoothness parameter are allowed. By performing a joint analysis, it is 
found that $\Omega_m = 0.27^{+0.06}_{-0.03}$ and $\alpha \geq 0.52$.  As a general result, although considering that current GRB data alone cannot constrain the smoothness $\alpha$ parameter  our analysis provides an interesting 
cosmological probe for dark energy even in the presence of inhomogeneities.   
\end{abstract}

\maketitle

\section{Introduction}

It is widely known that the Friedman-Lemaitre-Robertson-Walker (FLRW) uniform geometry 
provides a very successful description of the Universe at large scales ($\ell \geq 100$ Mpc). 
However,  due to the structure formation process, the inhomogeneities present at small and
moderate scales influence the trajectories of
light beams thereby producing observable phenomena like the ones
associated to gravitational lensing. Since lensing effects must cause either brightening or dimming of 
cosmic sources, a basic consequence of the inhomogeneities is to alter the  cosmic distances in 
comparison to the standard homogeneous description. In other words, any cosmic distance 
calculated along the line of sight (l.o.s) of the local observers must be somehow corrected by 
taking into account the presence of inhomogeneities. 

At present, the solution of the problem related to the light propagation in the framework 
of a late time clumpy Universe is far from a 
consensus \cite{Mattsson10,B11}. One possibility to deal
with the inhomogeneities is to consider them in randomly distributed
compact clumps with higher density compensated by a lower density of the 
smoothly distributed matter.  The distance
obtained in such an approach is called Dyer-Roeder distance \cite{Dy72,SFE92},
although its necessity was already discussed by Zeldovich
\cite{Ze64} and Kantowski \cite{Kant69}. Then we refer to it here as the
Zeldovich-Kantowski-Dyer-Roeder (ZKDR) distance (for an overview on
cosmic distances taking into account the presence of inhomogeneities
see the paper by Kantowski \cite{Kant03}). In this model, the effects experienced by the light beam is phenomenologically quantified by the
smoothness $\alpha$ parameter.  There are two limiting cases, namely: $\alpha =
1$ (filled beam), where the FLRW uniform distances are fully recovered and $\alpha = 0$ (empty beam) 
which represents the limit of a totally clumped universe. Therefore, for a
partial clumpiness, the smoothness parameter lies on the interval $0 <
\alpha < 1$.  Notice that in this model only demagnification happens. This is physically expected by the fact that 
light travels preferentially in voids, with light in denser environments being absorbed or scattered.

There is a rich literature concerning the ZKDR approach and its applications to cosmology.  
Investigations involving many different physical aspects 
and phenomenologies were performed, among them: analytical expressions
\cite{Kant98,Kant00,DEM03}, critical redshift for the angular
diameter distance ($d_A$) \cite{SPS01,LIB02}, i.~e., the redshift where
$d_A$ attains its maximum value, time delays \cite{GA01,LIB02}, gravitational
lensing \cite{koc02,koc03} and accelerating Universe models driven by particle
creation \cite{CdS04}. More recently, some quantitative analysis  by using ultra-compact radio sources as
standard rulers \cite{AL04,SL07} and type Ia supernovae as standard candles 
\cite{SCL08} were also performed.

In a previous analysis, Santos {\it et al.} \cite{SCL08} applied the
ZKDR approach for a flat $\Lambda$CDM model by considering two different samples
of SNe Ia, namely, the Astier {\it et al.} (2006) sample
\cite{Astier06} and the gold sample of Riess {\it et al.} (2007)
\cite{Ries07}. The first sample, composed by low redshifts
supernovae, provided no constraints to the smoothness parameter,
while the latter, which is composed by higher redshifts supernovae,
restricted it over the interval $0.42 \leq \alpha \leq 1.0$. In principle, such a result is strongly suggesting  
that  objects at  redshifts higher than those probed by SNe Ia could constrain the
smoothness parameter. Since Gamma-ray Bursts (GRBs) have been detected up to 
redshifts $z \sim 8$, they are the natural candidates to test such a conjecture. 

So far, GRBs are the most luminous explosions observed in the
Universe. Recent theoretical and observational developments have
shown that the presence of  afterglows and that their best candidates to progenitors are 
core-colapse supernovae (for 
comprehensive reviews on GRB physics see Refs. \cite{piran04,Meszaros06}). 
Still more important, the possibility of 
applying them as standard candles has also been
discussed by several authors \cite{ghirlanda2006,schaefer}. Recently, some
studies employing GRBs have shown that they may provide a complementary
test to constrain cosmological parameters
\cite{Costa2011,schaefer,standardgrb,standardgrb1}. Indeed, GRBs are also very promising tools for cosmology
from many different viewpoints. In particular, the association of long GRBs with peculiar type Ib/c SNe or
hypernovae, and thus the death of very massive stars, is supported both by theories and observations \cite{Woosley}. 
Thus, given their huge luminosity and redshift distribution extending up to at least
$z \approx 8$, GRBs may be considered powerful and unique tracers for the evolution of  the star formation rate up to the
reionization epoch \cite{Salvaterra, Petitjean}.

In this paper, by assuming a flat
$\Lambda$CDM model, we derive new constraints to the smoothness parameter
$\alpha$ and the matter density parameter $\Omega_m$. The ZKDR inhomogeneous distance approach it will be adopted here, however,  
different from \cite{SCL08}, the present statistical analysis it will be based on the 557 SNe Ia from Union2 Compilation 
Data \cite{Union2} plus 59 Hymnium GRBs \cite{HaoGRB}. 
As we shall see, the current SNe Ia and GRBs samples separately used do not provide tight constraints to 
the $\alpha$ parameter. Nevertheless, our joint
analysis restricts the pair of parameters ($\Omega_m, \alpha$) on the intervals $0.24 \leq \Omega_m \leq 0.33$ and $0.52
\leq \alpha \leq 1.0$ within $95.4$\% confidence level ($2\sigma$).
As an extra bonus, it is also found that the Einstein-de Sitter model is excluded with high
statistical confidence level, and, as such, our analysis provides an interesting cosmological probe for dark energy even in 
the presence of inhomogeneities.  

The paper is organized as follows. In section II, we present the
basic equations and the distance description by taking into account the
inhomogeneities as described by the ZKDR equation. In section III,
we determine the constraints on the cosmic parameters from the SNe
Ia and GRBs samples either separately and also through a joint analysis involving both samples. 
Finally, we summarize the main conclusions in section IV.

\begin{figure*}
\centerline{\epsfig{figure=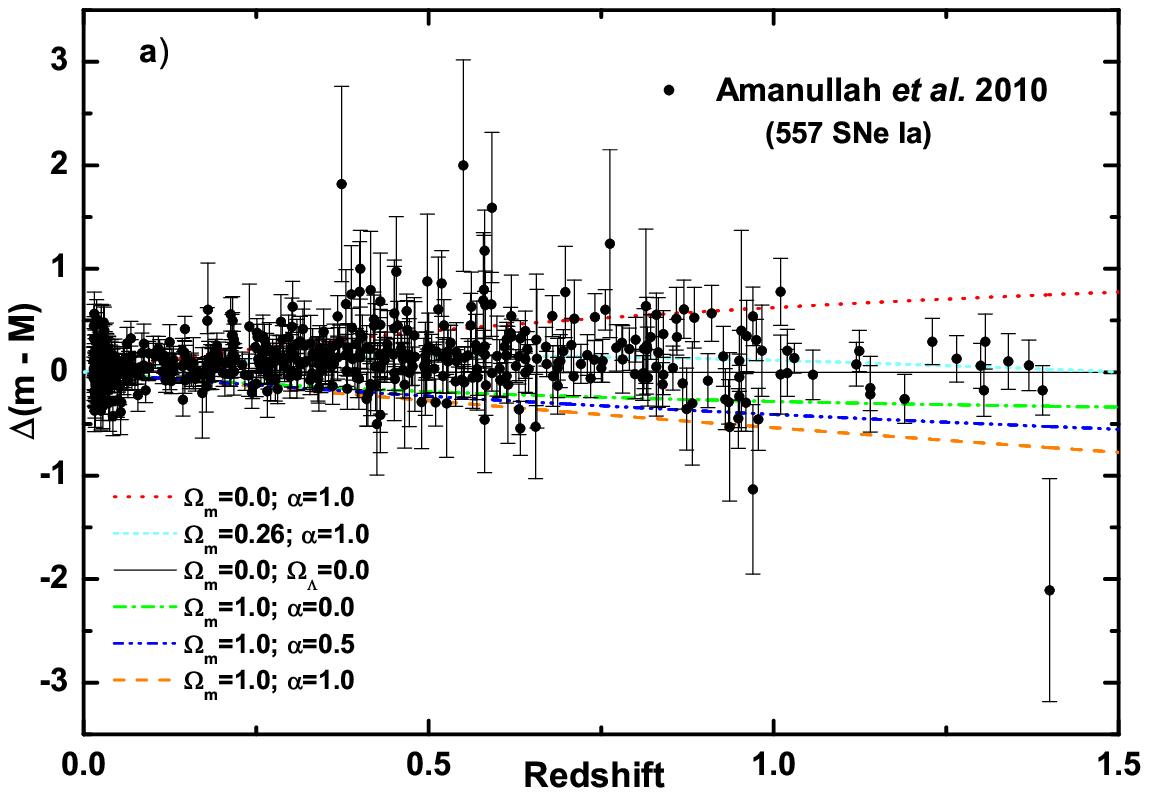,width=3.1truein,height=2.5truein}
\epsfig{figure=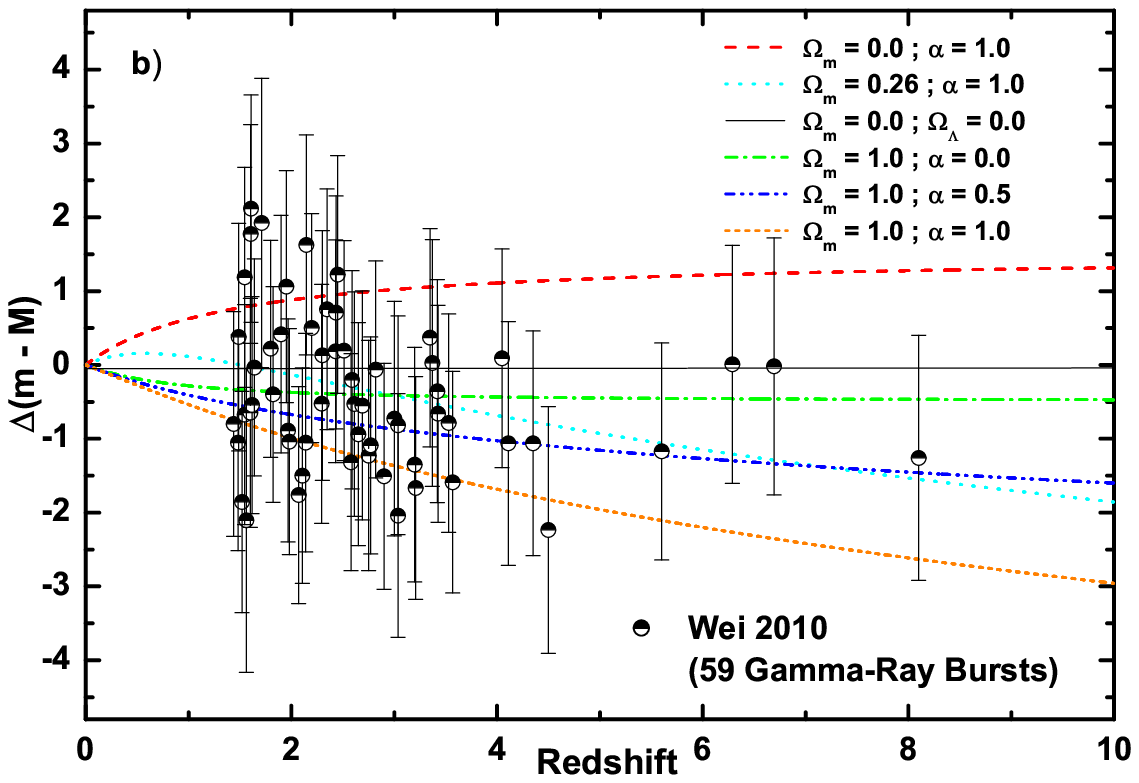,width=3.1truein,height=2.5truein}
\hskip 0.1in} \caption{The $\alpha$-effect on the residual magnitudes. {\bf a}) The 557 Supernovae data from the Union2 Compilation Data \cite{Union2} and the predictions of the ZKDR luminosity distance for several 
values of $\alpha$ relative to an empty model of the Universe ($\Omega_m=0$ and $\Omega_{\Lambda}=0$). {\bf b}) The same graph but now for the  59 Hymnium GRBs \cite{HaoGRB}. For comparison, in both panels we see  
the predictions (light blue curves) of the cosmic concordance model ($\Omega_m=0.26$, $\Omega_{\Lambda}=0.74$, $\alpha=1$).}
\end{figure*}

\section{ZKDR Equation for luminosity distance}

In order to describe the degree of inhomogeneity in the local
distribution of matter, we also adopt the phenomenological description based on the so-called
smoothness parameter, $\alpha$. This parameter was originally introduced by
Dyer \& Roeder \cite{Dy72}, when writing a
differential equation for the angular diameter distance in locally
clumpy cosmological models. To
obtain the ZKDR equation, let us consider the light propagation in
the geometric optics approximation \cite{Sachs61,SFE92}
\begin{eqnarray}\label{sachs}
{\sqrt{A}}'' +\frac{1}{2}R_{\mu \nu}k^{\mu}k^{\nu} \sqrt{A}=0,
\end{eqnarray}
where a prime denotes differentiation with respect to the affine
parameter $\lambda$, $A$ is the cross-sectional area of the light
beam, $R_{\mu\nu}$ the Ricci tensor, and $k^{\mu}$ the photon
four-momentum. In this form, it is implicit that the influence of
the Weyl tensor (shear) can be neglected. This means that the light
rays are propagating far from the  mass inhomogeneities  so that the
large-scale homogeneity implies that their shear contribution are
canceled \cite{shear}. Further, the Ricci tensor $R_{\mu\nu}$ is related to 
the energy momentum tensor 
$T_{\mu\nu}$ through the Einstein field equations: 
\begin{equation}
R_{\mu\nu}-\frac{1}{2} R g_{\mu\nu} = 8\pi G T_{\mu\nu},
\end{equation}
where $R$ is the scalar curvature, $g_{\mu\nu}$ is the metric
described by a FLRW geometry and $G$ is Newton's constant (in our units $c=1$). The
clustering phenomenon is introduced by considering the following
energy-momentum tensor ($\Lambda$CDM model)
\begin{equation}
T_{\mu\nu} = T_{\mu\nu}^{m} + T_{\mu\nu}^{\Lambda} = \alpha \rho_m u_{\mu} u_{\nu} + \rho_{\Lambda}g_{\mu\nu},
\end{equation}
where $u_{\mu}$ is the four-velocity of the comoving volume
elements, $\rho_m$ is the matter energy density, $\rho_\Lambda = \Lambda/8\pi G$, is the vacuum energy density  
associated to  the cosmological constant and
$\alpha=1-\frac{\rho_{cl}}{<\rho_m>}$ is the smoothness parameter
introduced by Dyer \& Roeder \cite{Dy72}. Such a parameter
quantifies the portion of matter in clumps ($\rho_{cl}$) relative to
the amount of background matter which is uniformly distributed
($\rho_m$). In general, due to the structure formation process, it
should be dependent on the redshift, as well as, on the direction
along the line of sight (see, for instance, \cite{SL07,Kasai} and
Refs. therein). However, in the majority  of the works $\alpha$ is
assumed to be a constant parameter.  

Now, we assume a flat $\Lambda$CDM cosmology, as well as the validity of the standard duality relation 
between the angular diameter and luminosity distances,
$d_L = (1+z)^2 d_A$, sometimes called the Etherington principle \cite{ETHER33}.  Since the 
the cross-sectional length
$A^{\frac{1}{2}}$ is proportional to the angular distance $d_A$, it is readily seen that Eq. (\ref{sachs}) can be rewritten 
for the dimensionless luminosity distance ($D_L=H_0d_L$) as \cite{Kant98,Kant00,DEM03,SPS01,GA01,SCL08}:

\begin{equation}\label{angdiamalpha}
 \left( 1+z\right) ^{2}{\cal{F}}
\frac{d^{2}D_L}{dz^{2}} - \left( 1+z\right) {\cal{G}}
\frac{dD_L}{dz} + {\cal{H}} D_L=0,
\end{equation}
which satisfies the boundary conditions:
\begin{equation}
\left\{
\begin{array}{c}
D_L\left( 0\right) =0, \\
\\
\frac{dD_L}{dz}|_{0}=1.
\end{array}
\right.
\end{equation}
This is the the ZKDR equation, where $\cal{F}$, $\cal{G}$ end $\cal{H}$ are functions of the cosmological parameters, expressed in terms of the redshift by:

\begin{eqnarray}
{\cal{F}}& =& \Omega_m + (1-\Omega_m )(1+z)^{-3},\nonumber
\\ \nonumber \\ {\cal{G}} &=& \frac{\Omega_m}{2}
+2(1-\Omega_m )(1+z)^{-3},\nonumber
\\ \nonumber
\\ {\cal{H}} &=& \left(\frac{3\alpha-2}{2}\right)\Omega_m
 + 2(1-\Omega_m)(1+z)^{-3}. \\
\nonumber
\end{eqnarray}
As remarked before, the $\alpha$ parameter appearing in the
$\cal{H}$ expression  quantifies the clustered fraction of the pressureless matter, and would 
be a redshift dependent quantity. Here we follow the standard treatment  
so that $\alpha$ is also assumed to be a constant (see, for instance, \cite{SCL08} and Refs. therein). 

\section{Samples and Results}

We know that the Universe is homogeneous only at large scales. Then,
a more realistic description is to consider that at 
moderate and small scales matter is clumped, being homogeneous only on average. As
light is affected by local quantities, not global, expressions like
the distance modulus $\mu(H_0,\Omega_m,\Lambda,z)$ must be altered
when the clumpiness phenomenon is taken into account through ZKDR equation.

\begin{figure*}
\centerline{\epsfig{figure=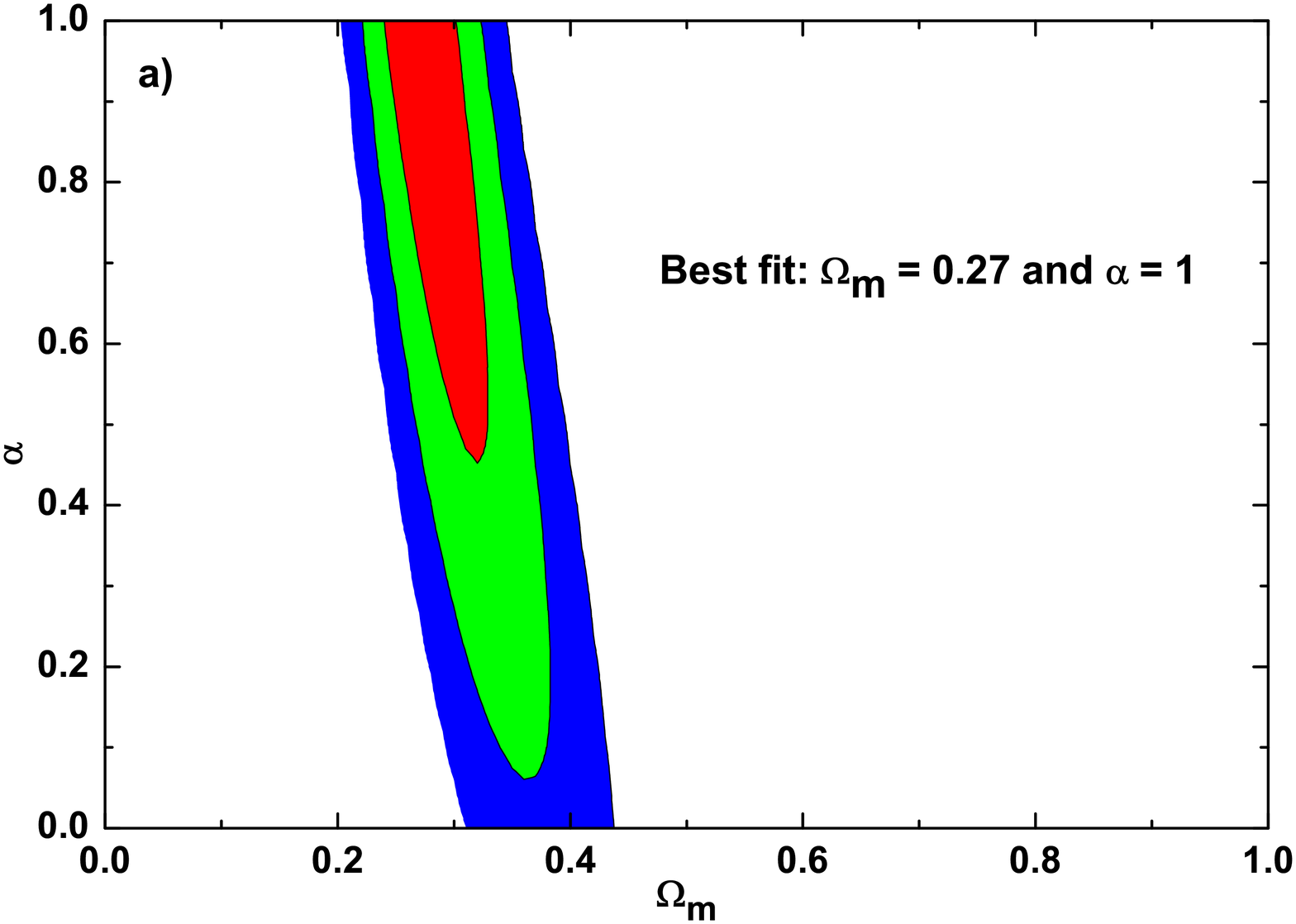,width=2.4truein,height=2.4truein}
\epsfig{figure=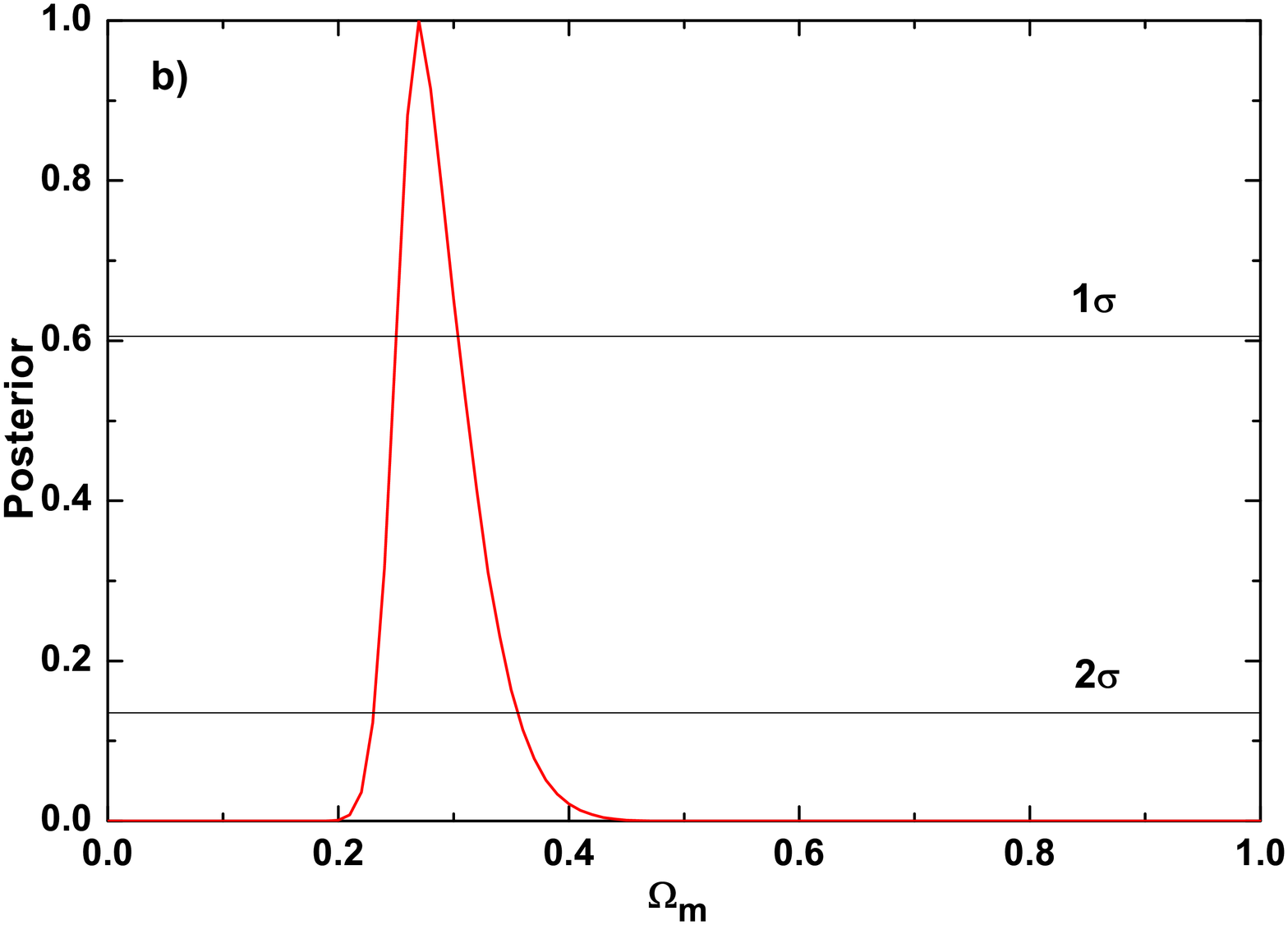,width=2.4truein,height=2.4truein}
\epsfig{figure=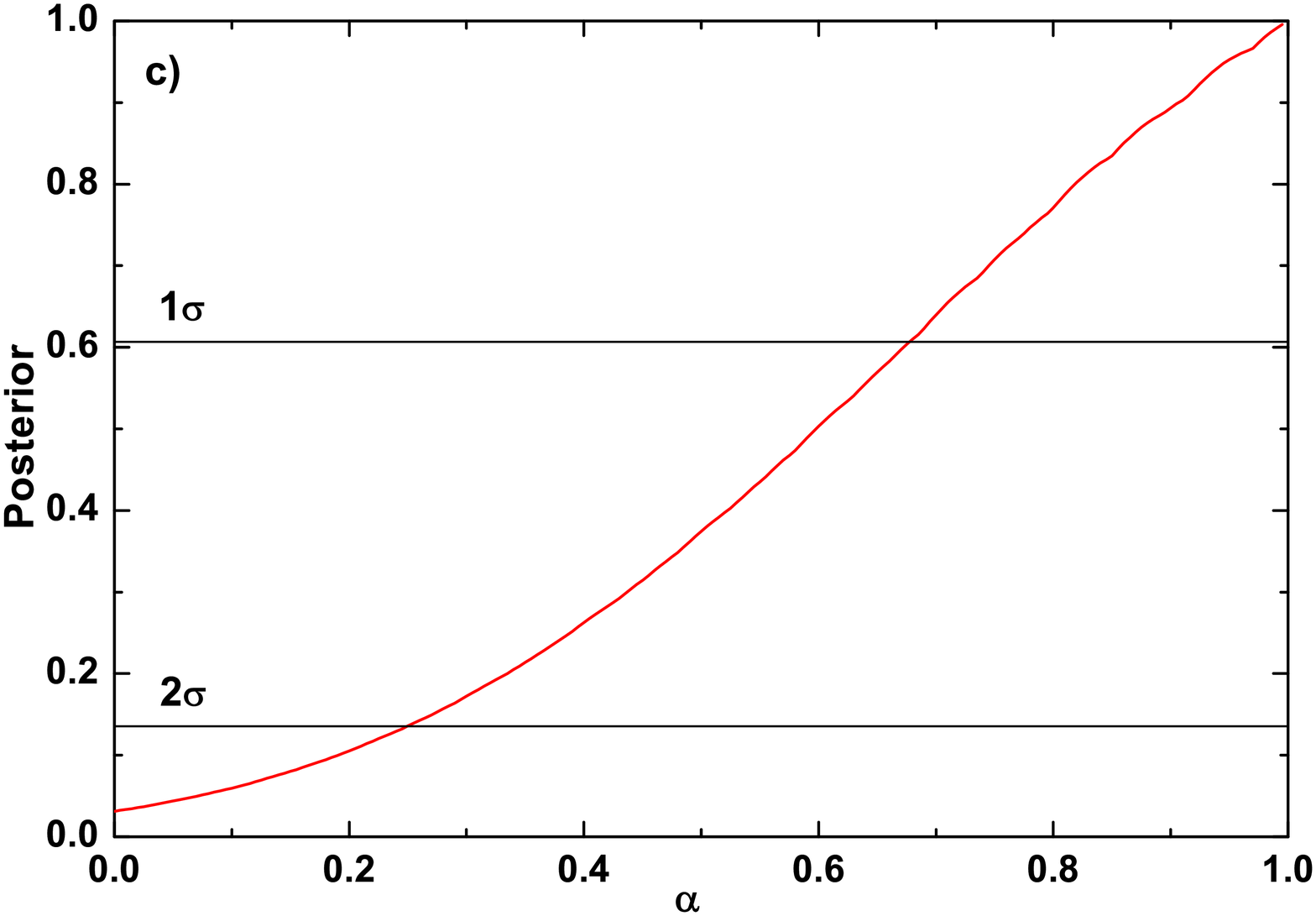,width=2.4truein,height=2.4truein}\hskip
0.1in} \caption{{\bf{a)}} The $\Omega_{m}-\alpha $ plane for flat
$\Lambda$CDM models obtained from 557 SNe Ia  from the Union2 Compilation Data \cite{Union2}. 
Contours stand for 68.3\%, 95.4\% and 99.7\% confidence levels. Note that the $\alpha$ parameter is not well
constrained by the data. {\bf{b)}} Posterior probability for the matter density parameter. We see that $0.24 \leq \Omega_m \leq 0.35$
with  $2\sigma$ confidence level. {\bf{c)}} Posterior probability
for the $\alpha$ smoothness parameter. We see that at 2$\sigma$ the
smoothness parameter is restricted on the interval ($0.25 \leq  \alpha \leq 1.0$).}
\end{figure*}

In Fig. 1, we display the effects of the inhomogeneities in the
reduced Hubble-Sandage Diagram for the Union2 \cite{Union2} and
Hymnium \cite{HaoGRB} samples for some selected values of the
smoothness parameter. The plots correspond to several values of
$\Omega_m$ and $\alpha$ as indicated in the panels. The difference
between the data and models from an empty universe case prediction
is also displayed there. For the sake of comparison, we also show
the Einstein-de Sitter (E-dS) model, i.e. $\Omega_m=1$ and $\alpha =
1$, as well as the present cosmic concordance ($\Omega_m=0.26$,
$\Omega_{\Lambda}=0.74$, $\alpha=1$). Note that cosmologies with only
matter and inhomogeneities can show a behavior resembling to some degree the cosmic
concordance model.

In order to constrain  the $\Omega_{\rm{m}}$ and $\alpha$ parameters,
a $\chi^{2}$ minimization will be applied for the sets of SNe Ia and
GRB data. Following standard lines, we maximize the posterior probability $\propto {\cal{L}} \times prior $ where :
\begin{equation}
{\cal{L}} \propto \exp(-\frac{\chi^{2}}{2}),
\end{equation}
we adopt a gaussian prior for the nuisance parameter $H_0$
centered at $74.2 \pm 3.6$ ${\rm km}{\rm s}^{-1}{\rm Mpc}^{-1}$
\cite{riess2009}, which will be marginalized, and $\chi^2$ is given by
\begin{equation}
\chi^2(z|\mathbf{p}) = \sum_i \frac{(\mu (z_i;
\mathbf{p})-\mu_{0,i})^2}{\sigma_{\mu_{0,i}}^2}.
\end{equation}

In the above expression $\mu$ is the theoretical distance modulus
for the set of parameters $\mathbf{p} \equiv (H_0,\alpha,\Omega_m)$,
$\mu_{0,i}$ is the measured distance modulus and
$\sigma_{\mu_{0,i}}$ its respective uncertainty. For a joint
analysis we just add the $\chi^2$ of each sample. We consider the
parameters $\alpha$ and $\Omega_{\rm{m}}$ restricted on the interval
[0,1] in steps of 0.01 for all numerical computations.

\subsection{SNe Ia}
Let us now discuss the bounds  arising from SNe Ia observations
on the  pair of parameters ($\Omega_{\rm{m}}, \alpha$) defining the ZKDR luminosity distance.

The Union2 Compilation Data \cite{Union2} are the largest 
SNe Ia sample and consist of 557 objects, where SALT2 light-curve fitter \cite{guy} was used to calibrate the supernovae events.
We have applied a $\chi^2$ minimization using this sample and the results are
displayed in Figs. 2a, 2b and 2c. We see from them that the
smoothness $\alpha$ parameter is poorly constrained, being restricted on the
interval $0.25 \leq \alpha \leq 1.0$ within $2\sigma$ confidence level.  
However, good constraints were obtained for the matter
density parameter, which is restricted on the interval $0.24 \leq
\Omega_m \leq 0.35 (2\sigma)$. Notice that a Universe composed only by
inhomogeneously distributed matter ($\Omega_\Lambda = 0$) is
also strongly disfavored by these data.

At this point, it is convenient to  compare the results derived here with a previous analysis performed by
Santos {\it et al.} \cite{SCL08} using 182 SNe Ia from the {\it
gold} sample observed by Riess {\it et al.} \cite{Ries07}. Within
$2\sigma$ c.l. the following intervals were achieved: $0.42 \leq \alpha
\leq 1.0$ and $0.25 \leq \Omega_m \leq 0.44$. It is interesting that the increasement in
the number of supernovae data provides a better constraint to
$\Omega_m$, but a greater range for the smoothness parameter is now 
allowed. We can understand this behavior by noting that low redshift
data are compatible with a more inhomogeneous set of data.  Indeed, this fact is in agreement with the conclusion that the structure 
formation process leads to a more locally inhomogeneous Universe thus, with a greater sample, it is more
likely to detect the effects caused by the inhomogeneities. Further, since the smoothness 
parameter appears only in third order in the $D_{L}$(z) expansion \cite{Mattsson10}, it is
interesting to investigate the parameter space ($\Omega_{\rm{m}},
\alpha$) using higher redshifts data. This will be examined in the next section.
\begin{figure*}
\centerline{\epsfig{figure=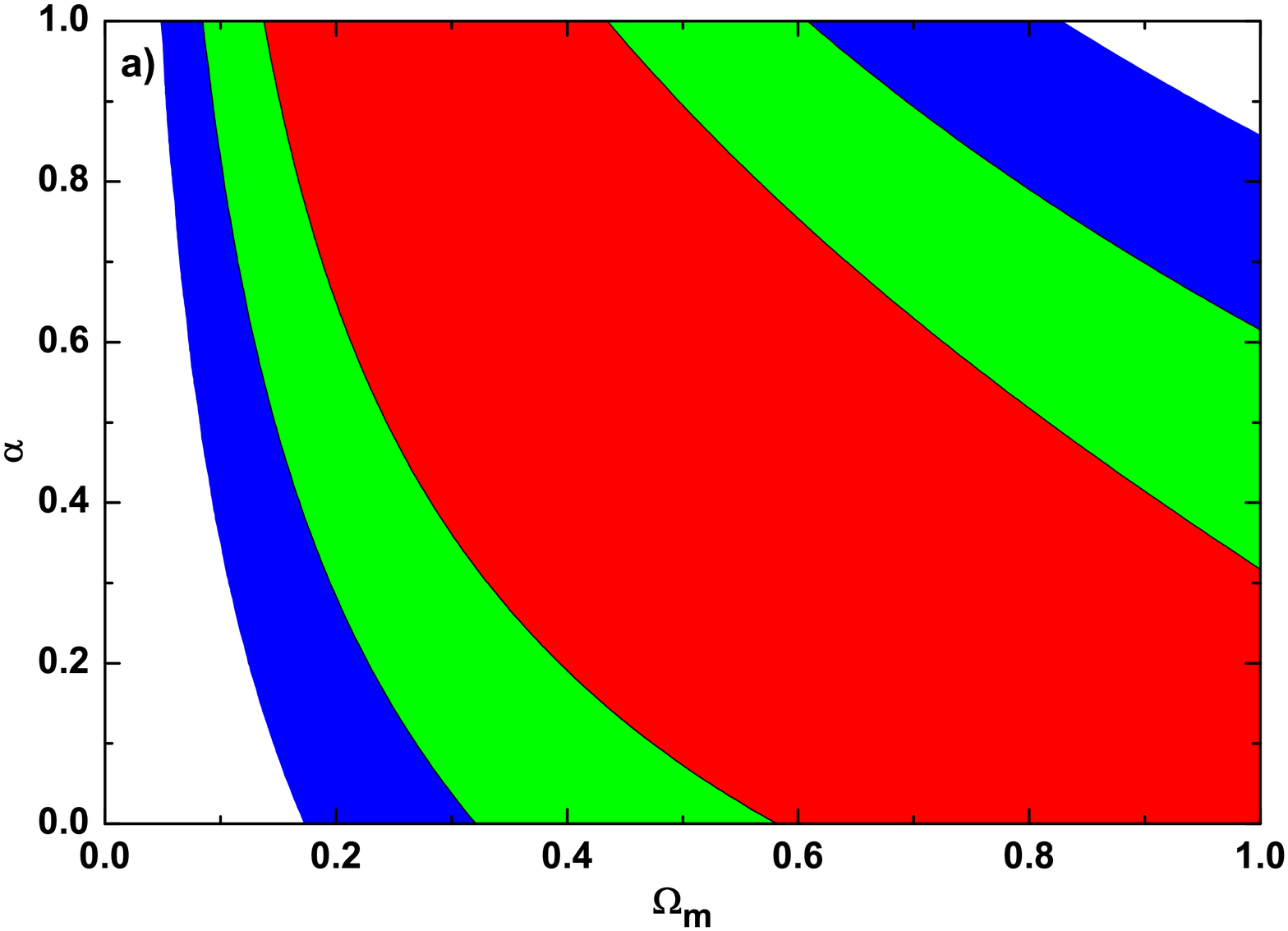,width=2.4truein,height=2.4truein}
\epsfig{figure=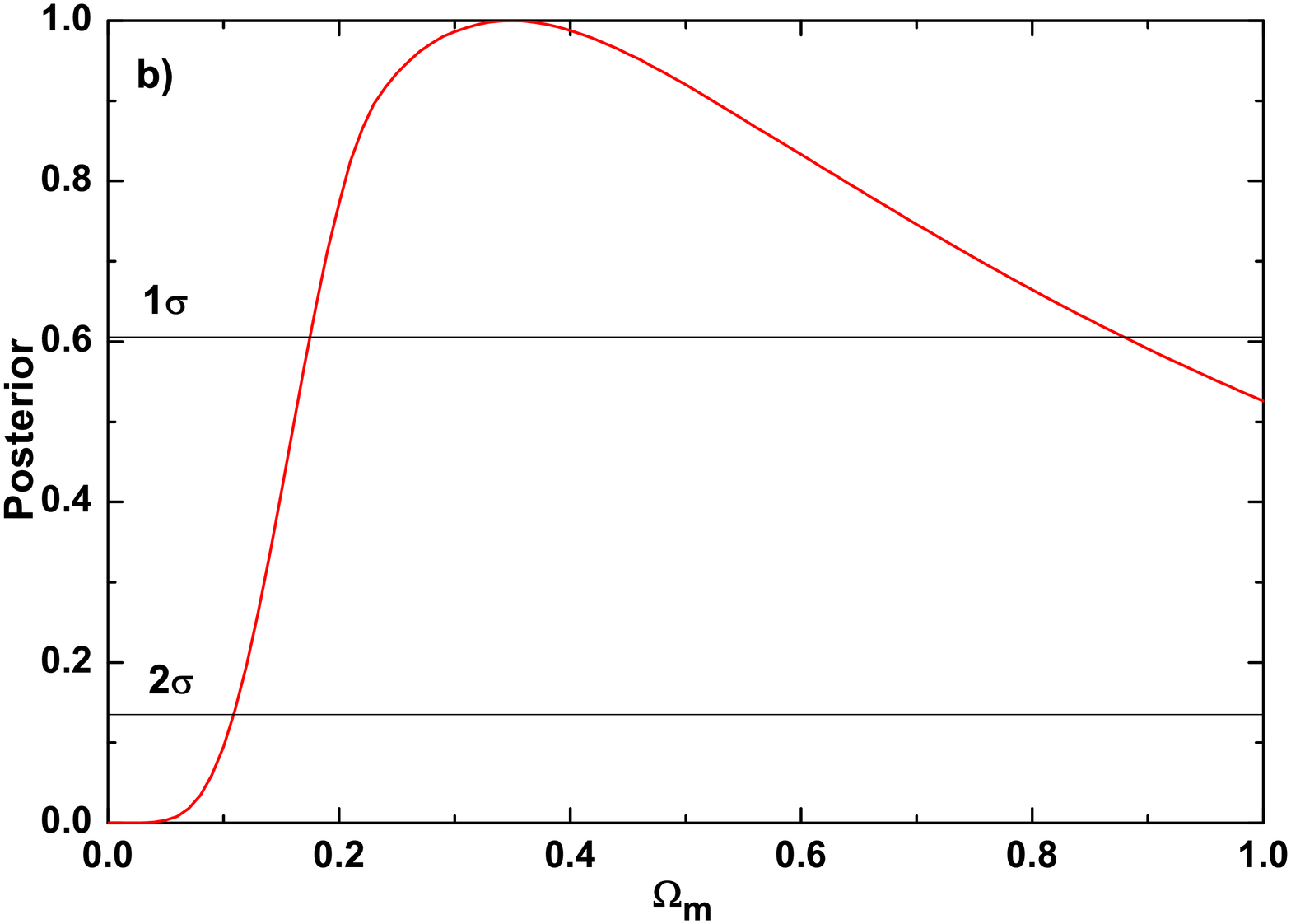,width=2.4truein,height=2.4truein}
\epsfig{figure=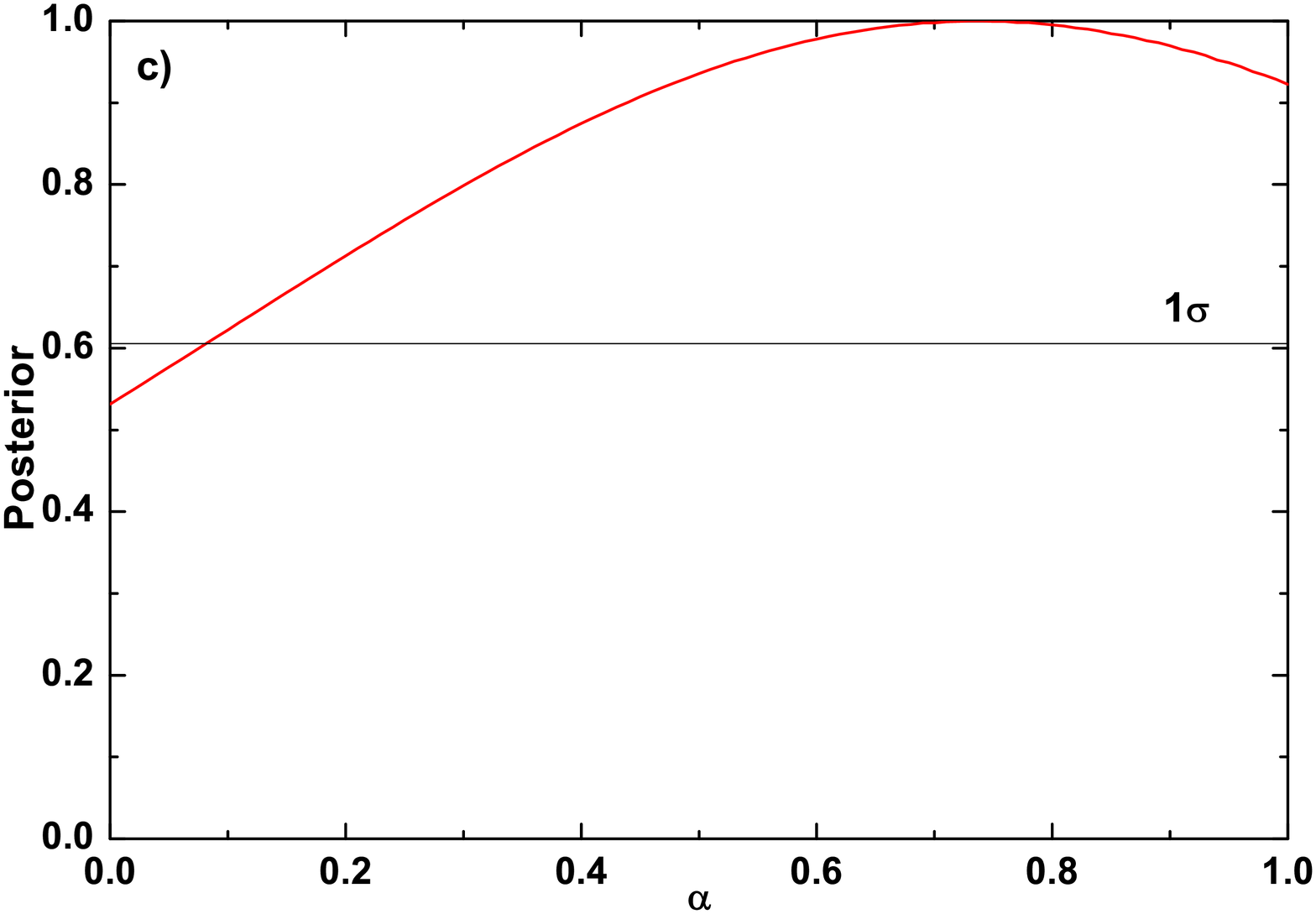,width=2.4truein,height=2.4truein}
\hskip 0.1in} \caption{{\bf{a)}} Contours of 68.3\%, 95.4\% and 99.7\% confidence on the ($\Omega_m,\alpha$) 
plane for  flat $\Lambda$CDM models as inferred from 59 Hymnium GRBs \cite{HaoGRB}. {\bf{b)}} Posterior probability 
for the matter density parameter. In this case almost all values are allowed within ($2\sigma$) confidence 
level ($0.11 \leq \Omega_m \leq 1.0$). {\bf{c)}}  Posterior probability for 
the $\alpha$ smoothness parameter. We see that at 2$\sigma$ the
smoothness parameter is not constrained by the data.}
\end{figure*}

\subsection{Gamma-Ray Bursts}

Gamma-ray bursts (GRBs) is now offering  a possible route  to probe the expansion history
of the Universe up to redshifts $z \sim 8$. However, it is widely known that before using GRBs to constrain cosmological models, their  
correlations must be firstly calibrated. Here we consider a relation between the isotropic-equivalent radiated energy in gamma-rays $(E_{iso})$ and the photon energy at which the $\nu F_\nu$ is brightest $(E_{peak})$, known as 
the Amati relation \cite{amati}.  This relation is a 
power law: $E_{p,i}=a \times E_{iso}^{b}$, where $E_{p,i}=E_{peak}\times (1+z)$ is the cosmological rest-frame spectral peak energy. The quantity $E_{iso}$ is defined  by:

\begin{equation}
E_{iso}= 4 \pi d_{L}^{2}S_{bolo}(1+z)^{-1},
\end{equation}
where $S_{bolo}$ is the bolometric fluence of gamma-rays in a given GRB and $d_L$ is its luminosity distance.

The general procedure to calibrate the relation for cosmological purposes is to use a low redshift sample, where the distance does not depend on the cosmological parameters. That is not the case for GRBs, since the observed nearby 
GRBs may be intrinsically different as GRB 980425 and GRB 031203 \cite{nearby}. So, the cosmological parameters one would like to constrain enter in the determination of the parameters of the correlation. This is called the circularity 
problem. Some attempts to overcome the problem have been studied in the literature \cite{circular,kodama,liang}. In this work, we use the method proposed independently by Kodama {\it et al.} \cite{kodama} and Liang {\it et al.} \cite{liang}, which was recently updated by Wei \cite{HaoGRB}. 

The method consists in using SNe Ia as a distance ladder to calibrate the GRBs. Since the distance moduli for the SNe Ia are known, a cubic interpolation is performed to determine the parameters $a$ and $b$ in the Amati relation for 
the low redshift GRBs ($z<1.4$). Then, the distance moduli for the highest GRBs are obtained and they can be used as standard candles without the circularity problem. In this connection, it is worth mentioning that the calibration 
of GRBs is still a quite controversial subject. Indeed, even the Amati relation has been contested by some authors (see, for instance, Ref. \cite{schaefer2011}).

Wei \cite{HaoGRB} used the 557 Union2 SNe Ia \cite{Union2} to calibrate 109 GRBs compiled in \cite{amatiamostra}. By applying a cubic interpolation with 50 low redshift GRBs ($z<1.4$), the parameters of the Amati relation were 
determined and the distance moduli for the other 59 GRBs were derived. This sample is called the Hymnium sample and can be used to derive cosmological parameters without the circularity problem. 

In Fig. 3 we display the results of our statistical analysis using the GRB Hymnium sample.  From Fig. 3a we see that both parameters are poorly constrained by the data. The likelihoods appearing in Figs. 3b and 3c allow us to 
get the following constraints within $2\sigma$: $\Omega_m=0.35^{+0.65}_{-0.24}$ while all values for the smoothness parameter are allowed within $2\sigma$. These data are also compatible with a model composed by inhomogeneously 
distributed matter ($\Omega_\Lambda=0$, $\Omega_m=1$). In principle, such a fact can be understood by noticing that the considered sample is dominated by data with high redshifts, and, therefore, just in a moment where the dark 
energy component does not play a prominent role for the cosmic evolution. Naturally, the low restriction over $\alpha$ may also reflect the poor quality of the current GRB data as seen by the intrinsic scatter in the Amati relation.
In this concern, although out of the scope of this work, it would be interesting to analyze how different phenomenological relations used to calibrate the GRBs can alter the current constraints.

\subsection{SNe Ia and Gamma-Ray Bursts}

It is widely recognized that joint analyses in cosmology usually provide a powerful tool to improve constraints in the basic cosmological parameters. Therefore, it is interesting to perform a statistical  analysis by 
combining the 557 SNe Ia from the Union2 Compilation Data \cite{Union2} with the 59 Hymnium GRBs \cite{HaoGRB}.

In Figs. 4a, 4b and 4c we display the main results of our joint analysis. As can be seen from Fig. 4a, 
 the constraints on both parameters are  significantly  improved. The best fit
obtained is $\Omega_m=0.27$ and $\alpha=1$, with a
$\chi^{2}_{min}=568.36$. The confidence interval within $2\sigma$
for the matter density parameter was slightly changed ($0.24\leq
\Omega_m \leq 0.33$) compared to the SNe Ia sample while for the
smoothness parameter a great improvement was achieved ($0.52 \leq
\alpha \leq 1.0$) as compared to the limits individually obtained from each sample.  Again, the
Einstein-de Sitter model is excluded with high confidence.  The better restriction over $\alpha$ can be understood as follows: the high redshift 
GRB data  prefer a homogeneous Universe, and, as such, it should contribute to diminish the corresponding 
space parameter (see Fig. 4a). In other words, since the high redshift Universe
is more homogeneous, higher values of $\alpha$ are favored, exactly as happened.

In table I, we have summarized the main results of our joint analysis. 
\begin{figure*}
\centerline{\epsfig{figure=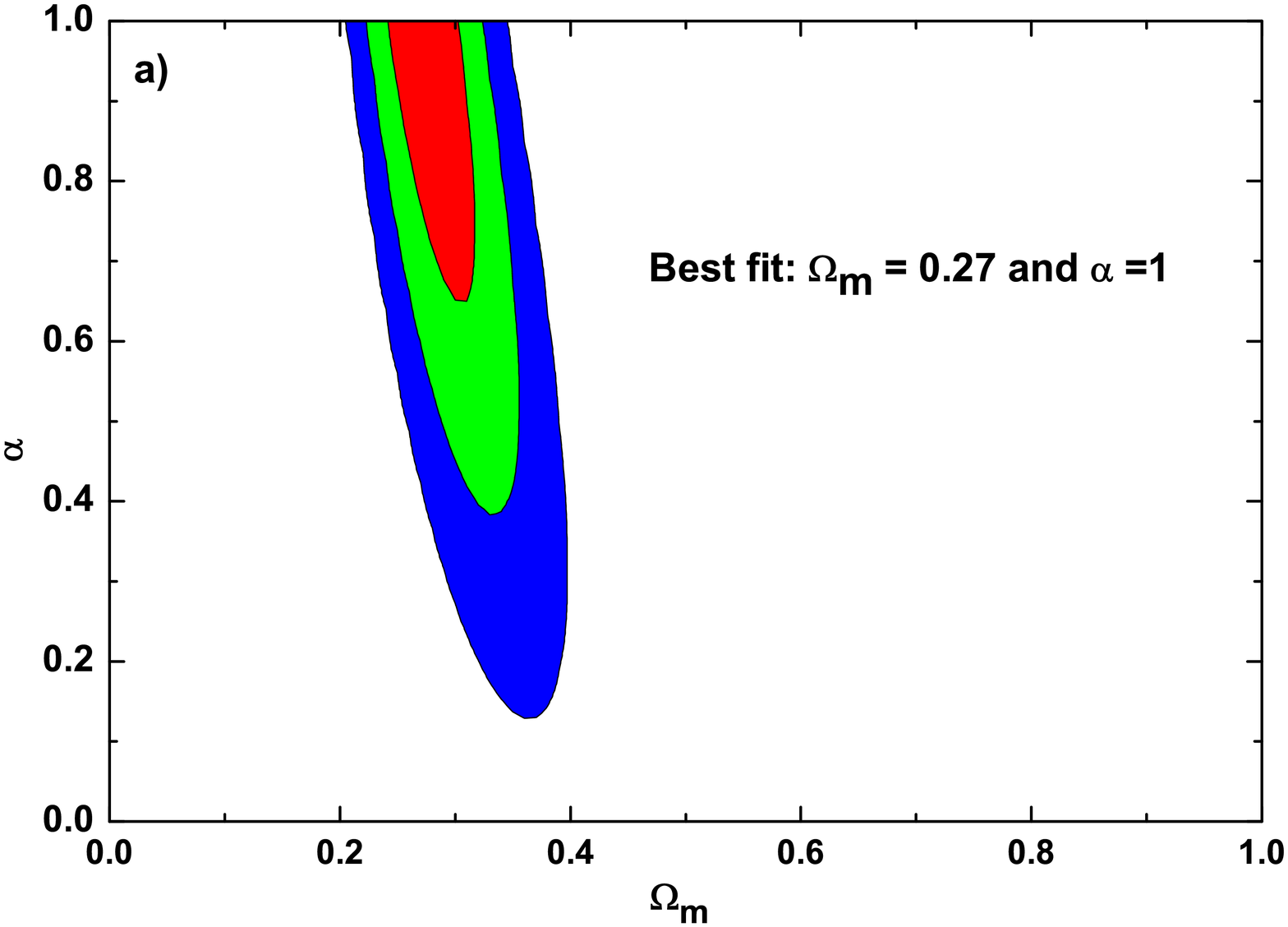,width=2.4truein,height=2.4truein}
\epsfig{figure=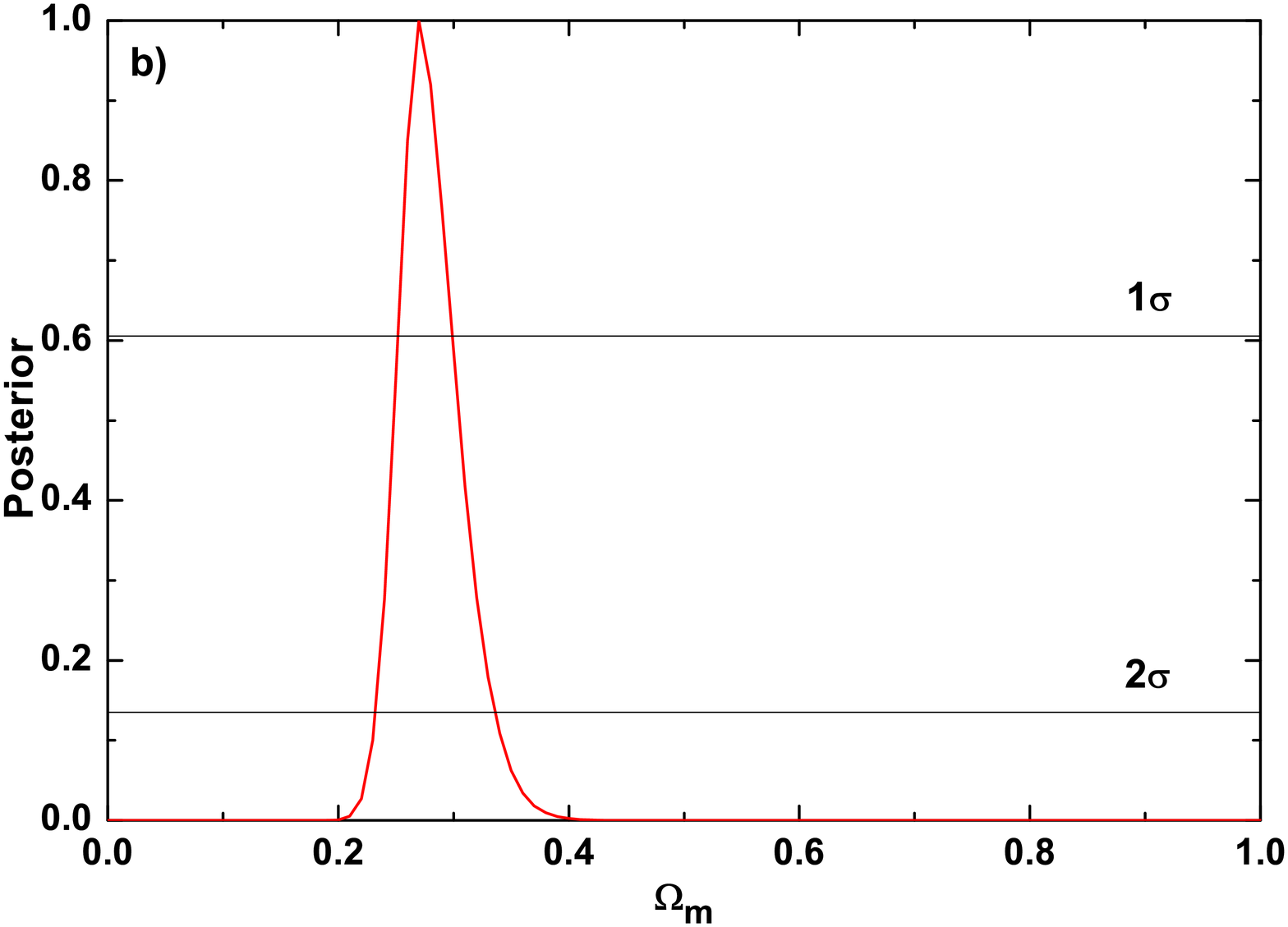,width=2.4truein,height=2.4truein}
\epsfig{figure=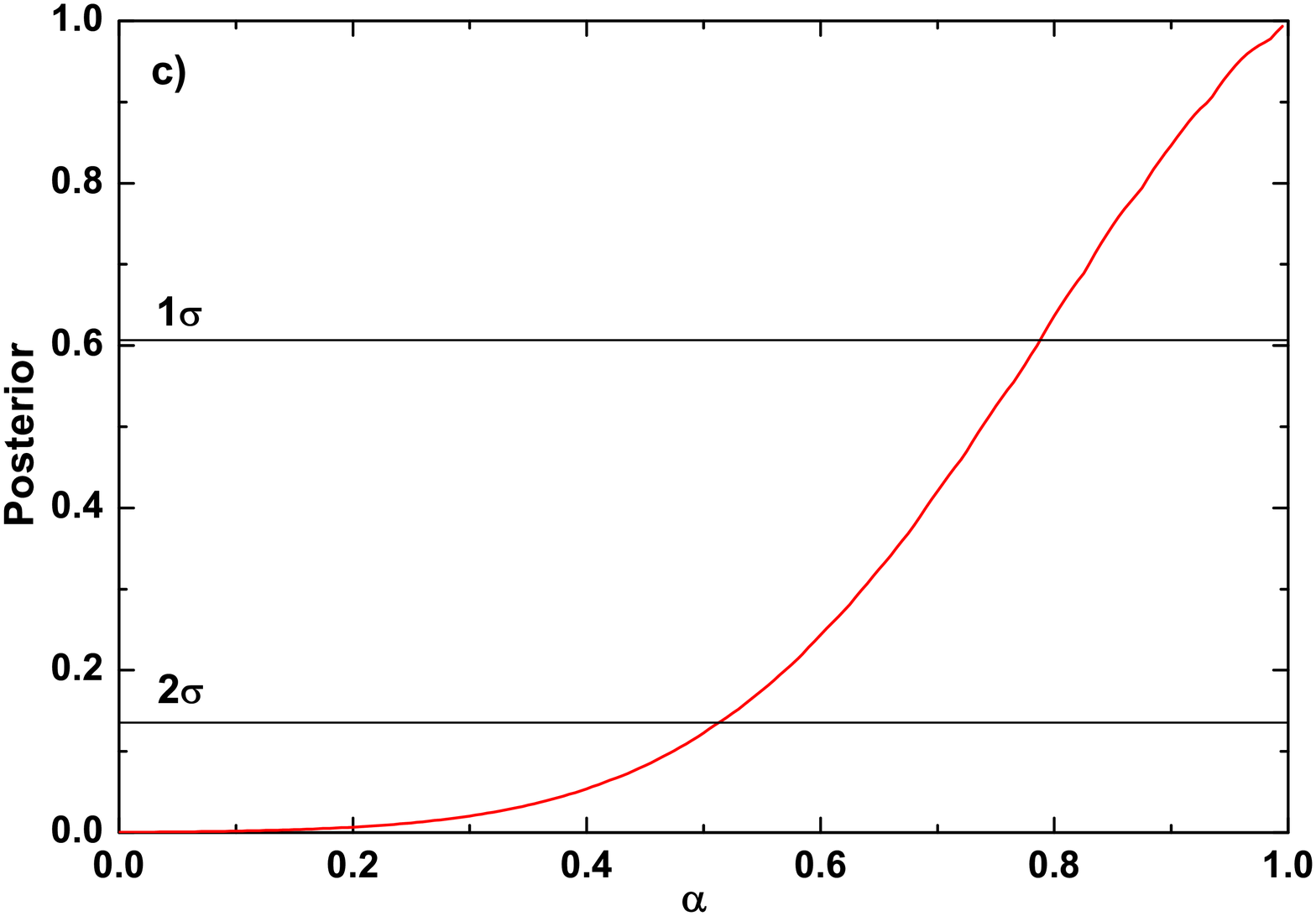,width=2.4truein,height=2.4truein}
\hskip 0.1in} \caption{{\bf{a)}} Contours of 68.3\%, 95.4\% and 99.7\% on the ($\Omega_m,\alpha$) 
plane for  flat $\Lambda$CDM models as inferred from 557 SNe Ia from the Union2 Compilation 
Data \cite{Union2} and 59 Hymnium GRBs \cite{HaoGRB}. {\bf{b)}} Posterior probability 
for the matter density parameter. In this case
a comparatively small region is permitted $0.24 \leq \Omega_m \leq 0.33$ with
($2\sigma$) confidence level.  {\bf{c)}} Posterior probability for the $\alpha$ 
smoothness parameter. We see that at 2$\sigma$ the
smoothness parameter is restricted on the interval ($0.52 \leq  \alpha \leq 1.0$).}
\end{figure*}

\begin{table}[htbp]
\caption{Limits to $\alpha$ and $\Omega_m$.}
\label{tables1}
\begin{center}
\begin{tabular}{@{}cccc@{}}
\hline Sample & $\Omega_m$ ($2\sigma$) & $\alpha$ ($2\sigma$) &
$\chi^2_{min}$
\\ \hline\hline
SNe Ia & $0.24 \leq \Omega_m \leq 0.35$ & $0.25 \leq \alpha \leq 1.0$&$545$ \\
GRBs &$0.11 \leq \Omega_m \leq 1.0$  & unconstrained     &$23$ \\
{\bf{Joint}}&{\boldmath{$0.24\leq \Omega_m \leq 0.33$}}&{\boldmath{$0.52\leq \alpha\leq 1.0$}}&{\boldmath{$568$}} \\
\hline
\end{tabular}
\end{center}
\end{table}

\section{Comments and Conclusions}

In the era of precision cosmology, it is expected that standard rulers and candles of
ever-increasing accuracy will provide powerful constraints on dark
energy and other cosmic parameters. However, in order to proceed with such a program it is also necessary  to
analyze carefully the physical hypotheses underlying the basic probes. It should also be recalled that even the large scale homogeneity 
(Copernican Principle) has been challenged in the last few years \cite{CP}. Besides,  we know that the Universe is effectively 
inhomogeneous at least in the small scale domain. In this concern,  the approach based on the ZKDR equation is a
simple alternative (together with weak lensing) for assessing quantitatively the effects of clumpiness phenomenon on the light propagation. 
As discussed here, it also provides a simple extension of the Hubble-Sandage  diagram thereby altering the standard cosmological
parameter estimation.

In this article, by using SNe Ia and GRBs samples, we have adopted the ZKDR approach to constrain the
influence of inhomogeneities in the context of a flat $\Lambda$CDM model. Our results are summarized in table I. 

We have shown that the SNe Ia sample \cite{Union2} was unable to constrain the
smoothness parameter, while the matter density parameter was well
constrained, being restricted on the interval $0.24 \leq \Omega_m
\leq 0.35 (2\sigma)$. Comparatively to a previous result
\cite{SCL08}, the smoothness parameter was less constrained even
with an increase of 375 supernovae. In principle, such a result may be justified based on the fact that  the Union2 sample has many low
redshift supernovae, this may suggest a redshift dependent
smoothness parameter already discussed by some authors
\cite{Kasai,SL07}. In addition, the GRBs sample also provided poor constraints for the
pair of parameters. In this case, the data are compatible with the present
Universe dominated only by inhomogeneously distributed matter
($\Omega_\Lambda=0$). This is also expected since at high redshifts dark
energy plays only a secondary role. The joint analysis provided good
constraints for both parameters. The intervals within $2\sigma$ were:
$0.52 \leq \alpha \leq 1.0$ and $0.24 \leq \Omega_m \leq 0.33$,
where the best fit was $\alpha=1.0$ and $\Omega_m=0.27$. 

It is important to point out that a smoothness parameter 
very different from unity is allowed by the current data, which may imply a cosmic concordance model with cosmological 
parameters shifted by several percent from the standard analysis. In particular, this means that the influence of the late time 
inhomogeneities can be important to decide  which is the best candidate to dark energy. For the near future, we believe that  new and more precise GRB data together 
the ZKDR approach (or some plausible extension of it) will play an important role in determining the real contribution of dark energy.

\vspace{1.0cm}

\noindent{\bf Acknowledgements:} The authors would like to thank J. V. Cunha, J. F. Jesus, R. F.
L. Holanda and F. A. Oliveira for helpful discussions. VCB is
supported by CNPq and JASL is partially supported by CNPq and FAPESP No. 04/13668-0
(Brazilian Research Agencies).  RCS is also grateful to INCT-Astrof\'isica and the Departamento de Astronomia (IAG-USP) 
for hospitality and facilities. 

\end{document}